%% file: prd_la.tex
\documentclass[twocolumn,showpacs,aps,prl,superscriptaddress]{revtex4}

\usepackage{graphicx}
\usepackage{dcolumn}
\usepackage{amsmath}
\usepackage{xspace}

\input pubboard/babarsym.tex

\input definitions.tex

\newcommand{\BABARPubYear}    {03}
\newcommand{\BABARPubNumber}  {026}

\newcommand{\SLACPubNumber} {10195}

\begin{document}

\preprint{\babar-PUB-\BABARPubYear/\BABARPubNumber}
\preprint{SLAC-PUB-\SLACPubNumber}

\begin{flushleft}
\babar-PUB-\BABARPubYear/\BABARPubNumber\\
SLAC-PUB-\SLACPubNumber\\
\end{flushleft}

\title{{\large {\bf Measurement of the Branching Fraction for {\boldmath{$\BchtoChiczeroKch$ }} }}}

\input pubboard/authors_jul2003.tex

\date{\today}

\begin{abstract}

We present a measurement of the branching fraction of the decay \mbox{$B^\pm \to \chi_{c0} K^\pm$} from a sample of 89 million \mbox{$B\kern 0.18em\overline{\kern -0.18em B}{}$} pairs collected by the \mbox{\slshape B\kern-0.1em{\smaller A}\kern-0.1em B\kern-0.1em{\smaller A\kern-0.2em R}} detector at the \mbox{PEP-II} asymmetric-energy $B$ Factory at SLAC. The $\chi_{c0}$ meson is reconstructed through its two-body decays to \mbox{$\pi^+ \pi^-$}  and \mbox{$K^+ \kern -0.16em K^-$}. We measure \mbox{$\cal B$$(B^\pm \to \chi_{c0} K^\pm)\times$$\cal B$$(\chi_{c0} \to \pi^+ \pi^-) = (1.32^{+0.28}_{-0.27}\mathrm{(stat)} \pm0.09\mathrm{(syst)})\times10^{-6}$} and \mbox{$\cal B$$(B^\pm \to \chi_{c0} K^\pm)\times$$\cal B$$(\chi_{c0} \to K^+ \kern -0.16em K^-) = (1.49^{+0.36}_{-0.34}\mathrm{(stat)} \pm0.11\mathrm{(syst)})\times 10^{-6}$}. Using the known values for the $\chi_{c0}$ decays branching fractions, we combine these results to obtain \mbox{$\cal B$$(B^\pm \to \chi_{c0} K^\pm) = (2.7\pm0.7)\times 10^{-4}$}.

\end{abstract}


\maketitle

In the simplest approximation, weak decays like $B \to \jpsi K$ arise from the quark-level process \mbox{$b \to \ccbar s$} through a current-current interaction that can be written \mbox{$[\cbar \gamma^\mu(1-\gamma_5)c][\sbar \gamma_\mu(1-\gamma_5)b]$}. 
The colorless current \mbox{$\cbar \gamma_\mu (1-\gamma_5)c$}, which can create the \jpsi, can create as well the $P$-wave state \chicone. It cannot, however, create \chiczero, \chictwo or $h_c$, so their appearance would have to be ascribed to more complex mechanisms.
The $b \to \ccbar s$ process also occurs through the interaction of two color-octet currents 
$J_8^{\mu (\cbar c)}J_{8 \mu}^{(\sbar b)} = \mbox{$[\cbar (\lambda_a/2) \gamma_\mu (1-\gamma_5)c]$}\mbox{$[\sbar (\lambda_a/2) \gamma^\mu(1-\gamma_5)b]$}$, where $\lambda_a$ are color $SU(3)$ matrices.
The current $J_8^{\mu (\cbar c)}$ can create a color-octet \ccbar\ pair in an $S$ state, which can then radiate a soft gluon to produce a $P$-wave bound state~\cite{ref:bodwin, ref:beneke}.  Alternatively, the \chiczero, \chictwo, $h_c$ states might arise from final state interactions that mix the $(\ccbar)K$ channel with channels like $D^{(*)} D_s^{(*)}$~\cite{ref:defazio}.

The first evidence for the \BchtoChiczeroKch\ decay was reported by the Belle collaboration~\cite{ref:belleChic0}, who measured $\BR(\BchtoChiczeroKch) = (6.0^{+2.1}_{-1.8} \pm 1.1)\times 10^{-4}$ on a sample of $31.3 \times 10^6$ \BB\ events. Previously CLEO had reported an upper limit of $\BR(\BchtoChiczeroKch) < 4.8\times 10^{-4}$ at \mbox{90$\%$ C.L.}~\cite{ref:cleoChic0}.

This work presents the study of the \BchtoChiczeroKch\ decay using data collected by the \babar\ detector operating at the \pep2\ asymmetric energy \epem\ collider. The data sample consists of 81.9 \invfb collected at the \FourS resonance containing $88.9 \times 10^6$ \BB pairs. 

The \babar\ detector is fully described elsewhere \cite{ref:babar}. It consists of a tracking system for the detection of charged particles, a Cherenkov detector (DIRC) for particle identification, an electromagnetic calorimeter and a detector for muon and \KL\ identification. The tracking system includes a 5-layer, double-sided silicon vertex tracker and a 40-layer drift chamber filled with a mixture of helium and isobutane, both in a 1.5-T magnetic field supplied by a superconducting solenoidal magnet. The DIRC is a novel imaging Cherenkov detector relying on total internal reflection in the radiator. The electromagnetic calorimeter consists of 6580 CsI(Tl) crystals. The iron flux return is segmented and instrumented with resistive plate chambers for muon and \KL\ identification.

Events with \BB\ pairs are selected by requiring the presence of at least three charged tracks, the ratio of the second to the zeroth order Fox-Wolfram moment~\cite{ref:FoxWolfram} to be less than 0.5 and the total energy of all the charged and neutral particles to be greater than 4.5 \gev. We consider only events where at least one track identified as a kaon has a momentum greater than 900 \mevc\ in the \epem\ center-of-mass frame.

We reconstruct the \chiczero\ meson in the decay modes \chiczeroToPipi\ and \chiczeroToKK\ from an oppositely-charged pair of tracks identified as both pions or both kaons, respectively. Candidates for the decay \BchtoChiczeroKch\ are formed by combining a track identified as a charged kaon (referred to as the ``bachelor'' kaon in the following) with a \chiczero\ candidate and performing a geometrical vertex fit. 
The efficiency for the kaon selection used is between $70\%$ and $90\%$, depending on momentum, while the probability for a pion to be misidentified as a kaon is below $5\%$. All the tracks are required to have polar angles in the region $0.35 < \theta < 2.54 \rad$, to have at least 12 hits in the drift chamber and a transverse momentum with respect to the beam direction larger than 100 \mevc. In addition, tracks consistent with being from \mbox{$\KS\to\pipi$}, \mbox{$\eta \to \pi^+\pi^-\pi^0$}, \mbox{$\Lambda \to p\pim$} (\mbox{$\Lbar \to \antiproton\pip$}) decays and tracks from \g\ conversions are rejected. The \chiczero\ candidates are required to have invariant mass in the range $3.32 < \mChiczero < 3.50 \gevcc$. 

To reject the large combinatorial background coming from continuum \qqbar\ events, a Fisher discriminant \fish~\cite{ref:fisher} is used, built from a linear combination of 11 quantities related to the event shape or the \B\ kinematics. The coefficients are determined by maximizing the separation between signal and continuum background on simulated events.
 
The selection of \B\ candidates relies on the kinematic constraints given by the \FourS\ initial state. Two variables are defined: the beam-energy substituted mass, \mbox{$\mes = \sqrt{(s/2 + \mathbf{p}_0 \cdot \mathbf{p})^{2}/E_0^2- |\mathbf{p}|^2}$}, and \mbox{$\DeltaE = E^* - \sqrt{s}/2$},
where $\mathbf{p}$ is the momentum of the \B\ candidate, $(E_0,\mathbf{p}_0)$ is the four-momentum of the initial state, in the laboratory frame, and $E^*$ is the \B\ candidate energy, $\sqrt{s}$ is the total energy, in the center-of-mass frame.
For the \BchtoChiczeroKch, \chiczeroToKK\ mode, when an ambiguity arises in cases with the same three final state kaons, we select as the bachelor kaon the one with the highest center-of-mass momentum.

The values of the cuts for \fish, \mes\ and \DeltaE\ are determined by an optimization procedure aimed at maximizing the value of $S/\sqrt{S+B}$. The number $S$ of signal candidates and $B$ of background events surviving the selection are estimated on samples of simulated events and data from the \DeltaE\ ``sidebands''  of the \mes-\DeltaE\ plane, respectively. The sidebands are defined by $5.2 < \mes < 5.3 \gevcc$, $0.1 < |\DeltaE| < 0.2 \gev$. The relative normalization of the signal and background samples is determined by assuming the value measured by Belle for \BR(\BchtoChiczeroKch) and the world average for \BR(\chiczeroToPipi) and \BR(\chiczeroToKK)~\cite{ref:pdg2002}. The signal regions in \DeltaE\ and \mes\ are defined by $-45 < \DeltaE < 35 \mev$, $\mes>5.2750 \gevcc$ for the \chiczeroToPipi\ mode and by $-70 < \DeltaE < 60 \mev$, $\mes>5.2735 \gevcc$ for the \chiczeroToKK\ mode. 

We apply a veto on fully reconstructed \mbox{$\B \to D^{(*)} h$} decays, where $h$ denotes a $\pi$, \kaon\ or $\rho$ meson. We reject \B\ candidates if at least one of their decay products also contributes to the reconstruction of a \mbox{$\B \to D^{(*)} h$} decay, with $|\DeltaE| < 30 \mev$ and $\mes>5.27 \gevcc$. To reduce the residual contamination from other \B\ decays with charmed or charmless mesons in the final state, we require the invariant mass of the pair formed by the bachelor kaon with the oppositely charged track from the \chiczero\ decay to be greater than \mbox{2 \gevcc}.

The main source of non-combinatorial background remaining after the selection described comes from non-resonant \B\ decays with the same final state as the signal, \mbox{$\Bpm \to \Kpm \pi^+ \pi^-$} and \mbox{$\Bpm \to \Kpm K^+ K^-$}. A reliable evaluation of the expected contamination from these processes cannot be obtained based on the available measurements. These modes are expected to behave as ``peaking background'', that is to peak in \mes\ and \DeltaE, while the distribution of \mChiczero\ is expected to be flat: this is used to separate their contribution from the signal by means of a fit to the data, as described below. 

The background from misreconstructed \chiczero\ decays to other modes is studied on simulated events and found to be negligible with respect to the other background sources for both the \pipi\ and the \KpKm\ modes.

The number of signal events is extracted by a simultaneous unbinned maximum likelihood fit to the \mes\ and \mChiczero\ distributions for the events in the \DeltaE\ signal band. Three components are assumed to contribute to the selected sample: a signal component, modeled with a non-relativistic Breit-Wigner function convolved with a Gaussian distribution in \mChiczero, and a Gaussian distribution in \mes; a combinatorial background component, modeled with a flat distribution in \mChiczero, and an Argus threshold function~\cite{ref:Argus} in \mes; a peaking background component, modeled with a flat distribution in \mChiczero, and a Gaussian distribution in \mes, assuming the same resolution as for the signal.

In the fit the \Bpm\ and \chiczero\ masses are fixed to their PDG values~\cite{ref:pdg2002}; the \chiczero\ width is fixed to the value recently measured by E835~\cite{ref:E835_chic0}, $\Gamma(\chiczero) = (9.8 \pm 1.0 \pm 0.1) \mevcc$. The width of the Gaussian peak in \mes\ and the \mChiczero\ resolution are determined from Monte Carlo samples. The Argus shape parameter and the relative weight of the three components are left as free parameters in the fit. 

We verify the goodness of the fit with the three-component model using a Monte Carlo technique. For each of the two \chiczero\ decay modes, we simulate a number of experiments by randomly generating samples of events distributed in \mes\ and \mChiczero\ according to the distributions used in the fit. The number of events generated for each sample is equal to the number of events in the corresponding real data sample; the parameters of the distributions are set to their fixed or fitted values. For each sample, the fit is repeated in the same conditions as on real data. The pulls for the number of signal and background events are distributed as expected. The probability of having a worse fit than the one on data is found to be about $65\%$ and $27\%$ for \chiczeroToPipi\ and \chiczeroToKK, respectively. 

We check the reliability of the yield extraction on a sample containing known amounts of combinatorial background, peaking background and signal events. We also verify the stability of the fit results against variations of the parameters fixed in the fit by floating them one at a time.

The signal and background yields resulting from the fit to the data are reported in Table~\ref{tab:yields}. The maximum correlation we observe is about $-40\%$, between the number of signal and peaking background events, for both the \pipi\ and the \KpKm\ modes. 
\begin{table}[!b]
\caption{Number of signal ($N_{\rm sig}$), combinatorial background ($N_{\rm comb}$) and peaking background ($N_{\rm pkg}$) events obtained by the fit described in the text (with statistical errors only).\\[0.5mm]}
\begin{ruledtabular}
\begin{tabular}{cccc} 
\rule[-1mm]{0mm}{1ex}Mode & $N_{\rm sig}$ & $N_{\rm comb}$ & $N_{\rm pkg}$ \\
\hline
\rule[-1mm]{0mm}{4ex}\chiczeroToPipi & $33.0^{+7.0}_{-6.7}$ & $111^{+12}_{-11}$ & $12.3^{+7.3}_{-6.3}$ \\
\rule[-1mm]{0mm}{4ex}\chiczeroToKK &  $30.4^{+7.3}_{-6.9}$ & $102^{+12}_{-11}$ & $22.2^{+8.5}_{-7.6}$ \\
\end{tabular}
\end{ruledtabular}
\label{tab:yields}
\end{table}
\begin{figure}[!hb]
\begin{center}
\includegraphics[scale=0.43]{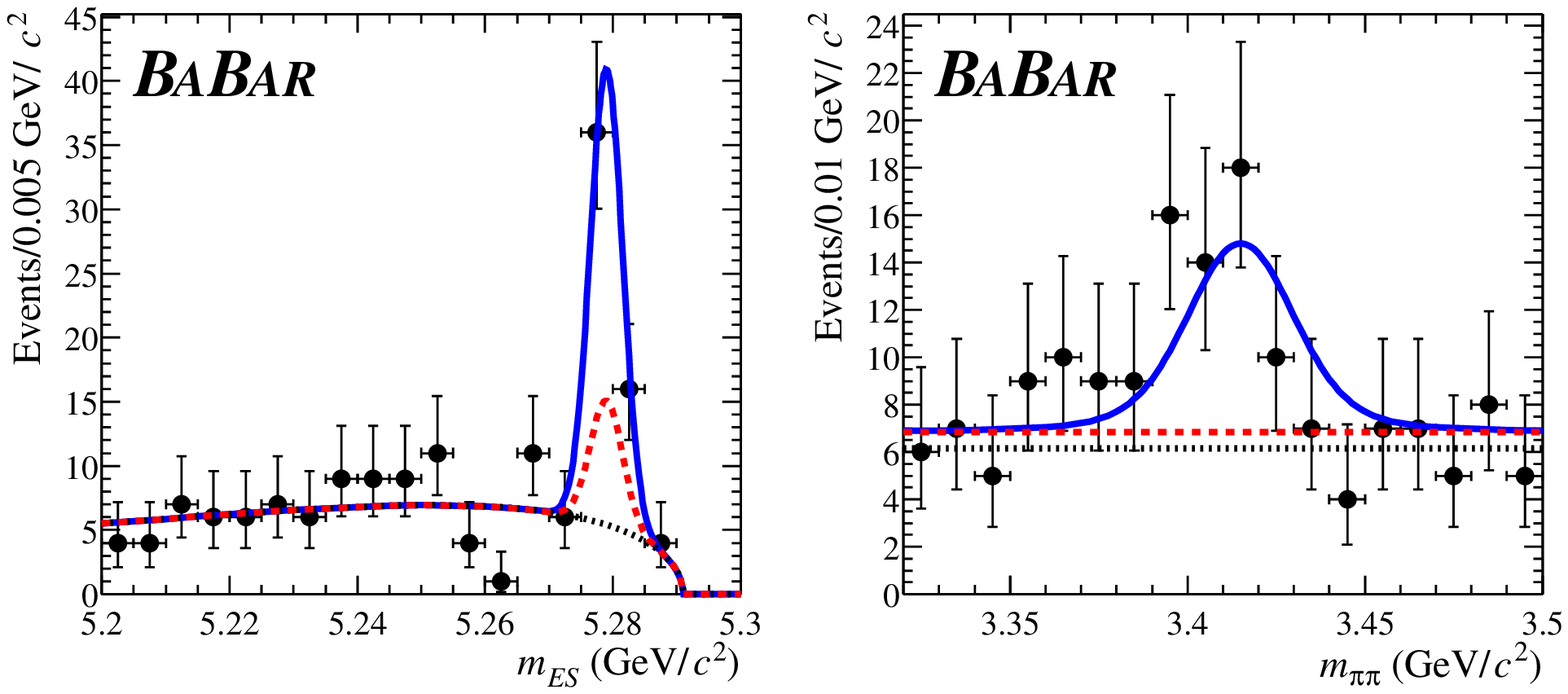}
\includegraphics[scale=0.43]{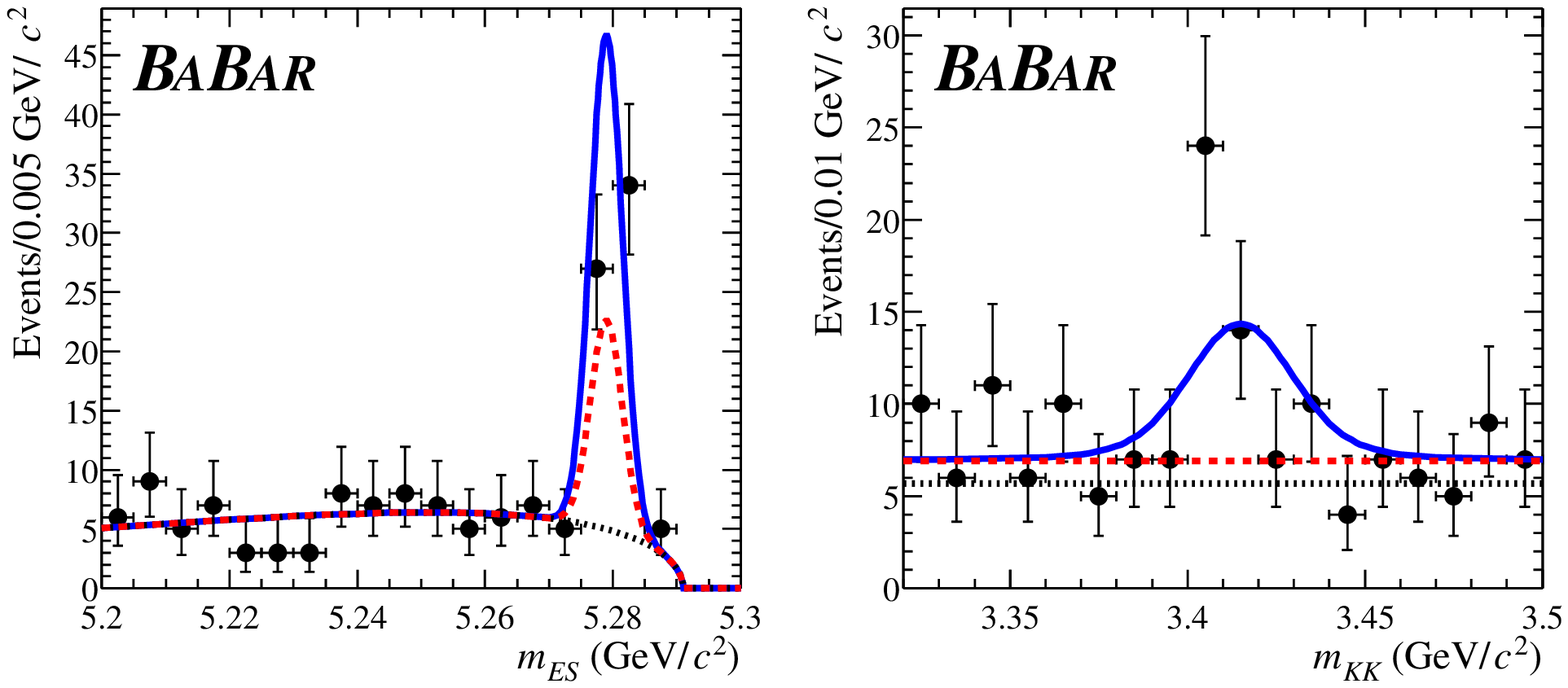}
\caption{Extraction of the signal yield for \chiczeroToPipi\ (top) and \chiczeroToKK\ (bottom). Left: \mes\ distribution; right: \mChiczero\ distribution. Dots with error bars represent the data; lines represent projection of the fitted functions for the three contributions: combinatorial background (dotted), peaking + combinatorial background (dashed), signal + backgrounds (solid).}
\label{fig:fit_pipi_KK}
\end{center}
\end{figure}
Figure \ref{fig:fit_pipi_KK} shows the \mes\ and the \mChiczero\ distributions for events in the \DeltaE\ signal region for the two modes considered. The results of the fit are superimposed.

We evaluate the systematic uncertainty to be attributed to the yield extraction by varying each fixed parameter, one at a time, by its error and repeating the fit. This results in a $2.4\%$ ($3.3\%$) fractional uncertainty for the \chiczeroToPipi\ (\chiczeroToKK) mode.

The statistical significance of the signal, defined as $\sqrt{2 \log{\cal L}_{max}/{\cal L}_0}$, where ${\cal L}_{max}/{\cal L}_0$ is the likelihood ratio for the fit with respect to the null signal hypothesis, is $8.1$ ($6.8$) standard deviations ($\sigma$) for the \chiczeroToPipi\ (\chiczeroToKK) mode. When the systematic uncertainties are taken into account, the significances of the signals become $7.7\sigma$ and $6.4\sigma$, respectively.

An alternative fitting method is employed to cross-check the results. An unbinned maximum likelihood fit to the \mChiczero\ distribution only is used to extract the yield for the events selected in the \mes-\DeltaE\ signal region. The signal component is modeled with a Breit-Wigner shape convolved with a Gaussian resolution function, while the background component is modeled with a linear function. This fit yields 
$N_{\rm sig}(\chiczeroToPipi) = 32.9^{+5.7}_{-6.1}$ and $N_{\rm sig}(\chiczeroToKK)  = 29.7^{+6.5}_{-6.6}$. Both values are compatible with the results obtained with the primary fitting method.

The overall selection efficiency, estimated by using simulated data, is $(27.4\pm1.5)\%$ for the \chiczeroToPipi\ mode and $(22.3\pm1.3)\%$ for the \chiczeroToKK\ mode. The quoted uncertainty is mainly due to differences between real and simulated data in the track reconstruction efficiency and in the efficiency of the \DeltaE\ cut for the \chiczeroToPipi\ mode and to differences between real and simulated data in the track reconstruction and particle identification efficiencies for the \chiczeroToKK\ mode.

We derive the branching fractions as \mbox{$\BR = N_{\rm sig}/(\epsilon N_{\Bpm})$}, where $\epsilon$ denotes the overall signal efficiency and $N_{\Bpm}$ is the total number of \Bpm\ mesons produced in the data sample considered. The value of $N_{\Bpm}$ is determined from the measured number of \BB\ pairs, \mbox{$N_{\BB} = (88.9\pm1.0) \times 10^6$}, and using \mbox{$\BR(\FourS \to \BpBm) = (0.513\pm 0.013)$}~\cite{ref:pdg2002}. 
We obtain:
\begin{equation}\nonumber
\BR ({\ensuremath{\Bpm \to \chiczero(\pipi) \Kpm}}) = 1.32^{+0.28}_{-0.27}\stat \pm0.09\syst,
\end{equation} 
\begin{equation}\nonumber
\BR ({\ensuremath{\Bpm \to \chiczero(\KpKm) \Kpm}}) = 1.49^{+0.36}_{-0.34}\stat \pm0.11\syst,
\end{equation} 
expressed in units of $10^{-6}$. The systematic error combines the uncertainties from the determination of the number of \BB\ pairs, from the branching fraction for $\FourS \to \BpBm$, from the yield extraction and from the signal efficiency. 

The ratio of the branching fractions for the \chiczero\ into the two modes is:
\begin{equation}\nonumber
\frac{\BR( \chiczero \to \pipi)}{\BR(\chiczero \to \KpKm)} = 0.88^{+0.28}_{-0.27}\stat \pm0.07\syst
\end{equation}
which is compatible within the quoted errors with the world average \cite{ref:pdg2002}.

\begin{table}[!t]
\caption{Branching fraction for \BchtoChiczeroKch\ with the two \chiczero\ decay modes. The meaning of the quoted errors and of the combined measurement is explained in the text.\\[0.5mm]}
\begin{ruledtabular}
\begin{tabular}{ll} 
Mode & $\BR(\BchtoChiczeroKch)\ (\times 10^{-4})$ \\
\hline
\rule[-2mm]{0mm}{4ex}\chiczeroToPipi & $2.83 ^{+0.60}_{-0.58} \pm 0.19 \pm 0.52$ \\
\rule[-2mm]{0mm}{4ex}\chiczeroToKK &  $2.63 ^{+0.63}_{-0.60} \pm 0.19 \pm 0.43$ \\
\hline
\rule[-1mm]{0mm}{4ex}combined & $2.7 \pm 0.7$ \\
\end{tabular}
\end{ruledtabular}
\label{tab:BR}
\end{table}
Using $\BR(\chiczero \to \pipi)=(4.68\pm 0.26\pm 0.65)\times 10^{-3}$ and $\BR(\chiczero \to \KpKm)=(5.68\pm 0.35\pm 0.85)\times 10^{-3}$, as reported by the BES collaboration~\cite{ref:BES_chic0}, we measure the values of \BR(\BchtoChiczeroKch) reported in Table~\ref{tab:BR}.
There the first quoted error is statistical, the second is systematic and the third accounts for the uncertainty on the \chiczero\ decay branching fractions. The measurements of $\BR (\BchtoChiczeroKch)$ obtained with the two \chiczero\ decay modes are compatible. Taking into account the correlated errors, we have combined them and derived the value reported in Table~\ref{tab:BR}.

In summary, we have studied the process \BchtoChiczeroKch, reconstructing the \chiczero\ meson through its decay modes \chiczeroToKK\ and \chiczeroToPipi; the measured branching fraction is $\BR (\BchtoChiczeroKch) = (2.7 \pm 0.7) \times 10^{-4}$.
The result is significantly different from the zero value expected from the color-singlet current-current contribution alone.

\input pubboard/acknow_PRL.tex

\end{document}

%% file: definitions.tex
\def\etactwos {\ensuremath{\etac(2S)}}

\def\BchtoChiczeroKch {\mbox{\ensuremath{\Bpm \to \chiczero \Kpm}}} 

\def\BchtoEtacKch {\ensuremath{\Bpm \to \etac \kern -0.18em \Kpm}} 
\def\BchtoJpsiKch {\ensuremath{\Bpm \to \jpsi \kern -0.18em \Kpm}} 
\def\BchtoEtactwosKch {\ensuremath{\Bpm \to \etactwos \kern -0.18em \Kpm}} 
\def\chiczeroToPipi {\mbox{\ensuremath{\chiczero \to \pipi}}}
\def\chiczeroToKK {\mbox{\ensuremath{\chiczero \to \KpKm}}}

\def\fish {\ensuremath{\cal F}}
\def\mChiczero {\ensuremath{m_{\chiczero}}}

%% file: pubboard/authors_jul2003.tex
%
\author{B.~Aubert}
\author{R.~Barate}
\author{D.~Boutigny}
\author{J.-M.~Gaillard}
\author{A.~Hicheur}
\author{Y.~Karyotakis}
\author{J.~P.~Lees}
\author{P.~Robbe}
\author{V.~Tisserand}
\author{A.~Zghiche}
\affiliation{Laboratoire de Physique des Particules, F-74941 Annecy-le-Vieux, France }
% ...
\collaboration{The \babar\ Collaboration}
\noaffiliation

%% file: pubboard/acknow_PRL.tex
We are grateful for the excellent luminosity and machine conditions
provided by our \pep2\ colleagues, 
and for the substantial dedicated effort from
the computing organizations that support \babar.
The collaborating institutions wish to thank 
SLAC for its support and kind hospitality. 
This work is supported by
DOE
and NSF (USA),
NSERC (Canada),
IHEP (China),
CEA and
CNRS-IN2P3
(France),
BMBF and DFG
(Germany),
INFN (Italy),
FOM (The Netherlands),
NFR (Norway),
MIST (Russia), and
PPARC (United Kingdom). 
Individuals have received support from the 
A.~P.~Sloan Foundation, 
Research Corporation,
and Alexander von Humboldt Foundation.